\documentclass[12pt,a4]{article}

\usepackage[utf8]{inputenc}
\usepackage[english]{babel}
\usepackage[T1]{fontenc}

\usepackage[titles]{tocloft}
\usepackage{authblk,lineno,hyperref,color,lscape,url,sectsty}
\usepackage[autostyle]{csquotes}

\usepackage[labelfont=bf]{caption}
\usepackage{ifpdf}
\usepackage{ulem}
\usepackage{soul}
\usepackage{soul}
\usepackage{textgreek}

\usepackage{float}
\usepackage{emptypage}

\usepackage{hyperref}
\usepackage{acronym}
\usepackage{ragged2e}
\usepackage{etoolbox}

\usepackage[nogroupskip,nonumberlist,acronym]{glossaries} 

\usepackage{setspace}
\usepackage{xcolor}

\usepackage[nottoc]{tocbibind}
\usepackage{graphicx}
\usepackage{caption}
\usepackage{subcaption}
\usepackage[a4paper,width=150mm,top=25mm,bottom=25mm]{geometry}
\usepackage{}
\usepackage{amsmath,amssymb,amsfonts,mathtools,notoccite}
\usepackage{chngcntr,rotating,pdflscape,enumitem,algpseudocode,algorithm,subcaption}
\usepackage{calligra}
\usepackage{array,float,multirow,graphicx,setspace,cite,url,titlesec,tocloft}
\usepackage{dashrule,arydshln}
\usepackage{fancyhdr}
\fancypagestyle{plain}{%
  \fancyhf{} %
  \fancyhead[C]{\footnotesize\itshape Preprint of an article published in Modern Physics Letters A (2026) 2650207. DOI: 10.1142/S021773232650207X}
  \fancyfoot[C]{\thepage} %
}

\title{Di-Higgs Production and Trilinear Higgs Self-Coupling in the scNMSSM: Qualitative Implications for the Electroweak Phase Transition at large $\lambda$ and Low $\tan\beta$}

\author{Marwa Telba\thanks{Corresponding author: \href{mailto:mrtelba@hotmail.com}{mrtelba@hotmail.com}}}

\date{}
\begin{document}

\maketitle
\section*{Abstract}
The trilinear Higgs self-coupling and di-Higgs production cross section are investigated in the semi-constrained NMSSM (scNMSSM) at large-$\lambda$ / low-$\tan\beta$ with full two-loop precision. A random scan of one million parameter points subject to theoretical, collider, flavor, and dark matter constraints yields 66 SM-like points with singlet fraction $S_{13}^2 < 0.05$, all found in the mass range $m_{h_1} \in [122.0,\,125.0]~\text{GeV}$ --- a feature that reflects an intrinsic property of the scNMSSM at large-$\lambda$ / low-$\tan\beta$, where doublet-dominated configurations are 
kinematically confined below ${\sim}125~\text{GeV}$. The trilinear coupling ratio is found to be universally suppressed: $\kappa_\lambda = 0.884\text{--}0.982$, $\langle\kappa_\lambda\rangle = 0.944 \pm 0.018$. The non-resonant di-Higgs cross section at $\sqrt{s} = 13.6~\text{TeV}$ lies in the range $30.3\text{--}36.9~\text{fb}$ ($0.88\text{--}1.05\,\sigma_{\rm SM}$), with the resonant contribution via $gg \to h_3 \to h_1h_1$ negligible ($\leq 0.004~\text{fb}$). All predictions are consistent with current ATLAS+CMS di-Higgs upper limits and is expected to be probed at the HL-LHC. These results are discussed in the context of the electroweak phase transition, where the universal suppression of $\kappa_\lambda$ is qualitatively consistent with a modified Higgs potential that could support a strengthened first-order transition. A quantitative determination of the phase transition strength via finite-temperature analysis is left for future work.

\section{Introduction}
The origin of the matter-antimatter asymmetry in the Universe remains one of the most profound mysteries in particle physics and cosmology \cite{Sakharov:1967dj}. Within the framework of the Standard Model (SM) of particle physics, the observed baryon asymmetry cannot be successfully generated due to the violation of the necessary Sakharov conditions \cite{Sakharov:1967dj}. This has motivated the exploration of beyond the Standard Model (BSM) scenarios that could provide a viable mechanism for baryogenesis, with electroweak baryogenesis being a particularly attractive possibility.

Electroweak baryogenesis relies on the departure from thermodynamic equilibrium during the electroweak phase transition (EWPT) in the early Universe, which can lead to the generation of a net baryon asymmetry \cite{Kuzmin:1985mm}. However, within the SM, the EWPT is a smooth crossover rather than a strong first-order phase transition, precluding the necessary out-of-equilibrium conditions for baryogenesis. Consequently, extensions of the SM that modify the Higgs sector and the dynamics of the EWPT have been extensively studied as potential avenues for successful electroweak baryogenesis.

The Next-to-Minimal Supersymmetric Standard Model (NMSSM) \cite{Ellwanger:2009dp} is a well-motivated extension of the Minimal Supersymmetric Standard Model (MSSM). By expanding the scalar sector with a complex singlet, the NMSSM provides a natural solution to the $\mu$-problem and significantly alters the Higgs potential dynamics compared to the MSSM. The presence of this singlet sector can facilitate a strong first-order electroweak phase transition (EWPT), satisfying one of the crucial Sakharov conditions for successful electroweak baryogenesis \cite{Huber:2006wf}.While various dimensions of Higgs phenomenology in supersymmetric scenarios have been explored—including constraints on scNMSSM bosonic decays from LHC data \cite{Telba:2020pji} and large deviations in trilinear couplings \cite{Baglio:2012np,Shao:2013bz}—a holistic understanding remains incomplete. Specifically, although previous studies have demonstrated that the NMSSM singlet sector alters di-Higgs production \cite{Ellwanger:2013ova,ellwanger2024benchmark} and can drive a strong first-order EWPT \cite{Huber:2006wf,Kozaczuk_2013}, these interconnected phenomena have not yet been analyzed simultaneously within the semi-constrained NMSSM (scNMSSM) framework. This lack of a combined analysis is particularly evident in the large-$\lambda$, low-$\tan\beta$ regime, which this work aims to address.

The trilinear Higgs self-coupling $\lambda_{hhh}$ plays a central role in connecting di-Higgs production phenomenology to the dynamics of the electroweak phase transition. In the SM, this coupling is uniquely fixed by the Higgs mass: $\lambda^{\text{SM}}_{hhh} = 3m_h^2/v = 191.0 \text{ GeV}$, corresponding to a second-order-like crossover transition that is insufficient for baryogenesis. In BSM scenarios where the EWPT is strengthened to first order, the Higgs potential is necessarily modified, and this modification manifests directly as a deviation of $\lambda_{hhh}$ from its SM value, parametrized by the ratio $\kappa_\lambda = \lambda_{hhh}/\lambda_{hhh}^{\text{SM}}$~\cite{Baglio:2012np,Shao:2013bz}. A measurement of $\kappa_\lambda \neq 1$ would therefore constitute direct evidence for a non-standard Higgs potential, with immediate implications for the viability of electroweak baryogenesis. 

In the NMSSM, the singlet-doublet mixing modifies the trilinear coupling through additional tree-level contributions proportional to $\lambda$, $\kappa$, and $\mu_{\text{eff}}$, making the large-$\lambda$ / low-$\tan\beta$ region particularly sensitive to these effects. This region is also the most natural from the perspective of the NMSSM Higgs mass: the tree-level contribution $\Delta m_h^2 \propto \lambda^2 v^2 \sin^2 2\beta$ is maximized at low $\tan\beta$, reducing the need for large radiative corrections and providing a more natural spectrum. The interplay between the naturalness of the Higgs mass, the modification of the trilinear coupling, and the implications for the EWPT in this region of the scNMSSM parameter space has not been previously studied in a systematic manner.

In this work, we present a systematic study of the trilinear Higgs self-coupling and di-Higgs production cross section in the semi-constrained NMSSM (scNMSSM) at large-$\lambda$ and low-$\tan\beta$, with GUT-scale boundary conditions and full current LHC constraints applied. We perform a random scan of one million parameter points using \texttt{NMSSMTools 6.1.2} with full two-loop Higgs mass precision, imposing theoretical, collider, flavor, and dark matter constraints. For the surviving SM-like points, we compute $\kappa_\lambda$ at tree level from the NMSSM mixing matrix and evaluate the non-resonant and resonant di-Higgs production cross sections at $\sqrt{s} = 13.6 \text{ TeV}$, projecting our sensitivity to future collider environments where applicable. Our results are compared with the SM prediction and with current LHC Run 3 limits, and their implications for the electroweak phase transition are discussed. 

The remainder of this paper is organized as follows. In Section 2, we outline the theoretical framework of the scNMSSM, focusing on the scalar Higgs potential and the trilinear couplings. Section 3 describes our numerical scanning methodology and the comprehensive theoretical and experimental constraints applied to the parameter space. Section 4 presents the core results of our numerical analysis, including the behavior of the trilinear Higgs self-coupling ratio and the corresponding resonant and non-resonant di-Higgs production cross sections. In Section 5, we discuss the cosmological implications of our findings for the electroweak phase transition. Finally, Section 6 provides our conclusions and future research directions.

\section{The scNMSSM Model}

The Higgs sector of the Next-to-Minimal Supersymmetric Standard Model (NMSSM) extends the Minimal Supersymmetric Standard Model (MSSM) by introducing a complex gauge-singlet superfield $\hat{S}$. The superpotential of the $\mathbb{Z}_3$-invariant NMSSM is given by:
\begin{equation}
W_{\text{NMSSM}} = W_{\text{MSSM}}|_{\mu=0} + \lambda \hat{S} \hat{H}_u \cdot \hat{H}_d + \frac{1}{3} \kappa \hat{S}^3
\end{equation}
where $\hat{H}_u$ and $\hat{H}_d$ are the MSSM Higgs doublet superfields, while $\lambda$ and $\kappa$ are dimensionless coupling parameters that govern the singlet-doublet interactions and the singlet self-interactions, respectively.

In addition to the superpotential, the soft SUSY-breaking Lagrangian contains terms that break supersymmetry softly. The soft breaking terms include the mass terms $m_{H_u}^2$, $m_{H_d}^2$, and $m_S^2$, as well as the trilinear coupling parameters $A_\lambda$ and $A_\kappa$:
\begin{equation}
-\mathcal{L}_{\text{soft}} = -\mathcal{L}_{\text{soft}}|_{\text{MSSM}} + m_S^2 |S|^2 + \left( \lambda A_\lambda S H_u \cdot H_d + \frac{1}{3} \kappa A_\kappa S^3 + \text{h.c.} \right)
\end{equation}

Upon electroweak symmetry breaking (EWSB), the scalar component of the singlet field $\hat{S}$ acquires a vacuum expectation value (VEV), $v_s = \langle S \rangle / \sqrt{2}$. This dynamically generates an effective $\mu$-term, defined as:
\begin{equation}
\mu_{\text{eff}} = \lambda v_s
\end{equation}
which naturally solves the MSSM $\mu$-problem at the electroweak scale. 

The Higgs doublet fields $H_u$ and $H_d$ expand around their respective vacuum expectation values $v_u$ and $v_d$, where the total electroweak VEV is defined in this convention as $v = \sqrt{v_u^2 + v_d^2} \approx 246.22\text{ GeV}$, and their ratio is parameterized as $\tan\beta = v_u/v_d$. The parameter space of the semi-constrained NMSSM (scNMSSM) is thus defined by specifying these parameters at the grand unification scale (GUT), ensuring that all experimental and theoretical requirements are successfully satisfied at the electroweak scale.

\section{The parameter space scanning strategy}

We began our investigation by performing a comprehensive scan of the scNMSSM parameter space using the public code $\mathtt{NMSSMTools}$ $\mathtt{ 6.1.2}$ \cite{NMSSMTools,Ellwanger:2004xm,Ellwanger:2005dv,Das:2011dg,Muhlleitner:2003vg}, taking into account all phenomenological constraints arising from Higgs searches in various channels at LEP and the LHC. The semi-constrained NMSSM is defined by universal soft supersymmetry-breaking terms at the GUT scale: a common scalar mass $m_{0}$, a common gaugino mass $M_{1/2}$, and a common trilinear coupling $A_{0}$. These are supplemented by the NMSSM-specific parameters $\lambda$, $\kappa$, $A_{\lambda}$, $A_{\kappa}$, and $\mu_{\text{eff}}$ specified at the SUSY scale. The boundary conditions are subsequently evolved from the grand unification scale (GUT) to the SUSY scale using two-loop renormalization group equations (RGEs).

To maximize the tree-level mass of the SM-like CP-even Higgs boson and minimize the fine-tuning associated with the electroweak scale, our numerical scan is strategically focused on the large-$\lambda$ and low-$\tan\beta$ regime. In this context, $\lambda$ is chosen to be sufficiently large to enhance the tree-level scalar mass via $\Delta m_h^2 \propto \lambda^2 v^2 \sin^2 2\beta$, yet bounded from above to preserve perturbativity up to the GUT scale. This simultaneously regulates the viable ranges for $\kappa$. Consequently, a random scan of one million parameter points was performed over the following specific boundaries:
\begin{align}
0.5 < \lambda < 0.7, \quad & 0.25 \le \kappa \le 0.5, \nonumber \\
2 \le \tan\beta \le 4, \quad & 100\text{ GeV} \le \mu_{\text{eff}} \le 160\text{ GeV}
\end{align}

The universal soft supersymmetry-breaking parameters at the GUT scale, along with the NMSSM-specific trilinear couplings, are scanned simultaneously to establish a phenomenologically viable sparticle spectrum. The remaining soft parameters are varied within the following intervals:
\begin{align}
-2500\text{ GeV} \le A_{0} \le -800\text{ GeV}, \nonumber \\
-1100\text{ GeV} \le A_{\lambda} \le -200\text{ GeV}, \nonumber \\
-1000\text{ GeV} \le A_{\kappa} \le 0\text{ GeV}, \nonumber \\
1000\text{ GeV} \le m_{0} \le 1700\text{ GeV}, \nonumber \\
500\text{ GeV} \le M_{1/2} \le 1100\text{ GeV}
\end{align}

The scan results are summarized in Table 1. The dominant rejection mechanism is the occurrence of tachyonic Higgs or sfermion masses, affecting 55.5\% of all points and reflecting the stringent requirements of radiative electroweak symmetry breaking with GUT-scale boundary conditions in the large-$\lambda$ regime. Constraint violations, primarily from LEP sparticle bounds and flavor observables, account for a further 30.6\% of rejected points, while RGE integration and convergence problems together account for the remaining 13.7\%. 

From the 2,343 surviving points, 80 satisfy the SM-like Higgs mass window $122\text{ GeV} \le m_{h_{1}} \le 128\text{ GeV}$. After additionally requiring that $h_{1}$ is doublet-dominated, defined by a singlet fraction component satisfying $S_{13}^{2} < 0.05$, 66 points constitute our final SM-like dataset. While the sample of 66 SM-like points is modest in size, it is representative of the accessible parameter space in the large-$\lambda$ / low-$\tan\beta$ region of the scNMSSM: the survival rate of 0.23\% reflects the stringent theoretical and experimental constraints rather than a limitation of the scan size. To verify that the universal suppression $\kappa_\lambda < 1$ is not an artifact of our selection criteria, we 
note that the 14 excluded points with $S_{13}^2 > 0.05$ also exhibit $\kappa_\lambda < 1$ (with values ranging down to $\kappa_\lambda \approx 0.15$ due to stronger singlet admixture), confirming that the suppression is a genuine feature of this parameter region and 
not a consequence of the SM-like selection cut. All 66 points were subsequently re-run individually with LHC Higgs coupling constraints enabled in \texttt{NMSSMTools}, and all passed without further reduction, confirming full compatibility with current LHC Higgs data.

\begin{table}[h]
\caption{Summary of the one-million-point scNMSSM random scan results.}
\label{points rejected}
\renewcommand{\arraystretch}{1.1} 
\centering
\begin{tabular}{|l|l|l|}
\hline\
Rejection criterion & Points rejected & Fraction \\
\hline
Tachyonic $m_{h_1}^2$, $m_{a_1}^2$, or $m_{h^\pm}^2$
& 534,235 & 53.4\% \\
Tachyonic sfermion masses & 20,555 & 2.1\% \\
Constraint violations & 305,991 & 30.6\% \\
RGE integration problems & 99,699 & 10.0\% \\
Convergence problems & 37,177 & 3.7\% \\
\hline
\textbf{Surviving points} & \textbf{2,343} & \textbf{0.23\%} \\
\hline
\end{tabular}
\label{tab:scan_results}
\end{table}

\section{Results and Collider Phenomenology}

In this section, we present the results of our numerical analysis 
for the 66 surviving SM-like scNMSSM points satisfying 
$S_{13}^2 < 0.05$ and $122~\text{GeV} \leq m_{h_1} \leq 128~\text{GeV}$. 
We focus on the behavior of the trilinear Higgs self-coupling ratio 
$\kappa_\lambda$ and the corresponding non-resonant and resonant 
di-Higgs production cross sections at the LHC with 
$\sqrt{s} = 13.6~\text{TeV}$.

\subsection{Higgs Mass Spectrum}

Of the one million generated points, 2,343 survived all constraints, 
of which 80 satisfy the SM-like Higgs mass window 
$122~\text{GeV} \leq m_{h_1} \leq 128~\text{GeV}$. 
The singlet fraction $S_{13}^2$ distribution of these 80 points is 
shown in Table~\ref{tab:singlet_frac}. The majority of points 
(66 out of 80, i.e.\ 82.5\%) have a small singlet component 
$S_{13}^2 < 0.05$, confirming that $h_1$ is predominantly 
doublet-like in this parameter region. The remaining 14 points 
have a significant singlet admixture and are excluded from the 
analysis of the SM-like Higgs properties. The threshold $S_{13}^2 < 0.05$ was chosen to ensure 
that $h_1$ behaves as a SM-like Higgs boson with a 
reduced coupling to gauge bosons satisfying 
$C_V = S_{11}\cos\beta + S_{12}\sin\beta > 0.975$, 
consistent with the LHC Higgs coupling measurements 
at the $2\sigma$ level~\cite{ATLAS:2015egz,
CMS:2014fzn}. To assess the robustness of our 
results with respect to this choice, we verified that 
relaxing the threshold to $S_{13}^2 < 0.10$ increases 
the sample size from 66 to 70 points, while the 
trilinear coupling ratio remains in the range 
$\kappa_\lambda \in [0.867,\,0.982]$, confirming that 
the universal suppression $\kappa_\lambda < 1$ is 
insensitive to the precise value of this cut.

The Higgs mass spectrum of the 66 SM-like points is summarized in Table~\ref{tab:param_ranges}. The lightest CP-even Higgs boson $h_1$ has mass 
$m_{h_1} \in [122,\,128]~\text{GeV}$, consistent with the observed SM-like Higgs boson at the LHC~\cite{ATLAS:2012yve, CMS:2012qbp} within the theoretical uncertainty of $\pm 2$--$3$ GeV inherent in the NMSSMTools two-loop Higgs mass 
calculation. The surviving points cluster in the range $122\text{--}125\text{ GeV}$ as shown in Fig.~\ref{fig:kappa_lambda}, reflecting the constraints of the large-$\lambda$ / low-$\tan\beta$ parameter space. Notably, all four points with $m_{h_1} > 125\text{ GeV}$ in the pre-selection sample of 80 points have a large singlet fraction $S_{13}^2 > 0.50$, indicating that $h_1$ is singlet-dominated in these cases and $h_2$ is the SM-like Higgs boson. This reveals an important structural feature of the scNMSSM at large-$\lambda$ / low-$\tan\beta$: a doublet-like $h_1$ consistent with SM Higgs measurements is confined to the mass range $122\text{--}125\text{ GeV}$, while higher masses are accessible only when $h_1$ acquires a large singlet admixture. This is a direct consequence of the NMSSM tree-level mass formula, where the $\lambda^2 v^2 \sin^2 2\beta$ contribution raises the doublet mass but simultaneously induces singlet-doublet mixing that prevents $h_1$ from being both SM-like and heavier than $\sim 125\text{ GeV}$ in this parameter region. The second CP-even state $h_2$ is heavier, with $m_{h_2} \in [131,\,268]~\text{GeV}$, ensuring that no resonant $h_2 \to h_1 h_1$ decay is kinematically accessible for any of the 66 points. The third CP-even state $h_3$ has mass $m_{h_3} \in [383,\,581]~\text{GeV}$ and is predominantly singlet-like, with a reduced top-quark coupling $|C_t(h_3)| \in [0.25,\,0.38]$ significantly below unity.

The lightest CP-odd state $a_1$ has mass 
$m_{a_1} \in [8,\,262]~\text{GeV}$, indicating the presence of a 
potentially light pseudoscalar in some regions of the parameter 
space. Points with $m_{a_1} < m_{h_1}/2$ are of particular 
phenomenological interest, as they open the decay channel 
$h_1 \to a_1 a_1$, which could modify the invisible width of $h_1$. 
However, the LHC Higgs coupling constraints applied in our scan 
ensure that the signal strengths of $h_1$ remain compatible with 
SM predictions within $2\sigma$, implicitly bounding any such 
contribution.

The singlet fraction of $h_1$, defined as $f_s = S_{13}^2$, 
lies in the range $[0.000,\,0.049]$ for our 66 SM-like points, 
with a mean value of $\langle f_s \rangle = 0.018$. This small but 
non-zero singlet admixture is responsible for the deviations of the 
Higgs couplings and the trilinear self-coupling from their SM values, 
as discussed in the following subsections.

\begin{table}[h]
\caption{Singlet fraction $S_{13}^2$ distribution of the 80 points 
with $122~\text{GeV} \leq m_{h_1} \leq 128~\text{GeV}$.}
\label{tab:singlet_frac}
\renewcommand{\arraystretch}{1.1}
\centering
\begin{tabular}{|l|l|l|}
\hline
$S_{13}^2$ range & Number of points \\
\hline
$0.00 - 0.01$ & 39 \\
$0.01 - 0.05$ & 27 \\
$0.05 - 0.10$ & 4  \\
$0.10 - 0.50$ & 6  \\
$> 0.50$      & 4  \\
\hline
\textbf{SM-like total} ($S_{13}^2 < 0.05$) & \textbf{66} \\
\hline
\end{tabular}
\end{table}

\subsection{The Trilinear Higgs Self-Coupling Suppression}

The trilinear Higgs self-coupling $\lambda_{h_1h_1h_1}$ is a crucial 
parameter for understanding the shape of the Higgs potential and its 
connection to the electroweak phase transition. It was computed at 
tree level for each of the 66 SM-like points using the NMSSM formula:

\begin{equation}
\lambda_{h_1h_1h_1} = \frac{3m_{h_1}^2}{v}\,C_V^3 
+ \frac{3\sqrt{2}\,\lambda\,\mu_{\rm eff}}{v}\,S_{13}\,C_V^2 
+ \frac{3\sqrt{2}\,\kappa\,\mu_{\rm eff}}{v}\,S_{13}^3
\label{eq:lambda_hhh}
\end{equation}

where $v = 246.22~\text{GeV}$ is the SM vacuum expectation value, 
and $C_V = S_{11}\cos\beta + S_{12}\sin\beta$ is the reduced coupling 
of $h_1$ to the $W$ and $Z$ bosons. The first term in 
Eq.~(\ref{eq:lambda_hhh}) is the SM-like doublet contribution, 
which dominates when $h_1$ is mostly doublet-like ($S_{13} \approx 0$). 
The second and third terms are NMSSM-specific corrections proportional 
to the singlet component $S_{13}$, arising from the $\lambda S H_u H_d$ 
interaction and the $\frac{\kappa}{3}S^3$ self-interaction of the 
singlet, respectively.

We quantify the deviation from the SM by defining the ratio:

\begin{equation}
\kappa_{\lambda} = \frac{\lambda_{h_1h_1h_1}}
{\lambda_{h_1h_1h_1}^{\rm SM}}, 
\qquad 
\lambda_{h_1h_1h_1}^{\rm SM} = \frac{3m_h^2}{v} = 191.0~\text{GeV}
\label{eq:kappa_lambda}
\end{equation}

where $m_h = 125.20~\text{GeV}$ is the observed SM Higgs mass.

\begin{figure}[h]
\centering
\includegraphics[width=\textwidth]{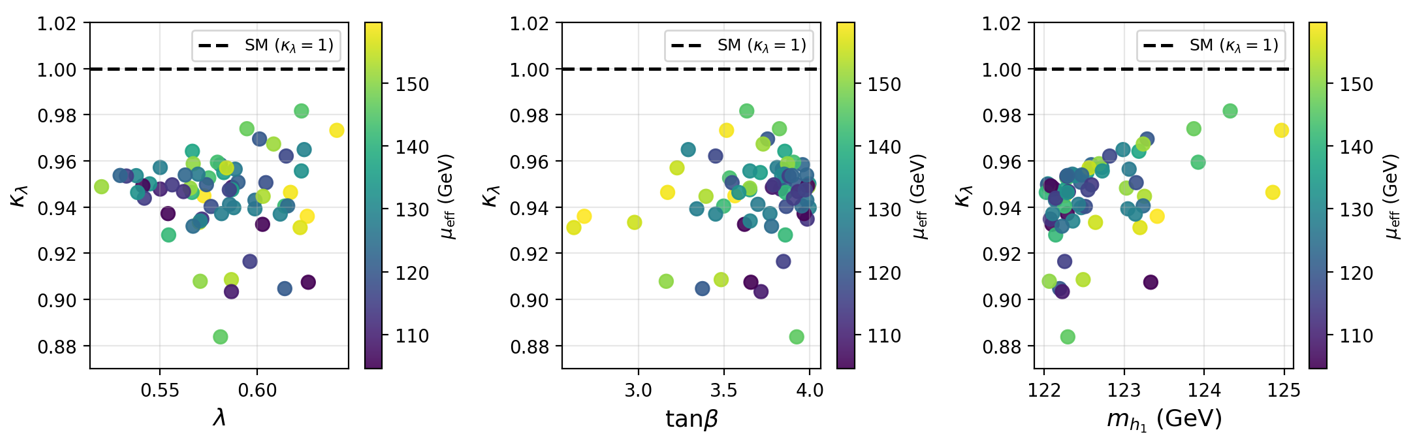}
\caption{Trilinear Higgs self-coupling ratio $\kappa_\lambda = \lambda_{h_1h_1h_1}/\lambda_{h_1h_1h_1}^{\rm SM}$ as a function of (left) the coupling $\lambda$, (center) $\tan\beta$, and (right) $m_{h_1}$, for the 66 SM-like scNMSSM points satisfying $S_{13}^2 < 0.05$ and and $122~\text{GeV} \leq m_{h_1}\leq 128~\text{GeV}$. The data points cluster in the range $122$--$125$ GeV due to the combined effect of the large-$\lambda$ constraint and the GUT-scale 
boundary conditions. The color coding indicates $\mu_{\rm eff}$ in GeV in all panels. The horizontal dashed line shows the SM prediction $\kappa_\lambda = 1$. All points lie below the SM value, demonstrating the universal suppression of the trilinear Higgs self-coupling in this parameter region.}
\label{fig:kappa_lambda}
\end{figure}

Figure~\ref{fig:kappa_lambda} displays the distribution of 
$\kappa_\lambda$ as a function of $\lambda$, $\tan\beta$, and 
$m_{h_1}$. Across all 66 validated points, the trilinear Higgs 
self-coupling is \textbf{universally suppressed} relative to the 
SM expectation, spanning a tight interval:

\begin{equation}
\kappa_{\lambda} = 0.884 - 0.982, 
\qquad 
\langle\kappa_{\lambda}\rangle = 0.944 \pm 0.018
\label{eq:kappa_result}
\end{equation}

This persistent suppression is primarily driven by the 
doublet-singlet scalar mixing characteristic of the 
scNMSSM in the large-$\lambda$ / low-$\tan\beta$ regime. 
Because $h_1$ retains a small but non-zero singlet 
fraction ($S_{13}^2 < 0.05$), the dominant term in 
Eq.~(\ref{eq:lambda_hhh}), which scales as $C_V^3$, 
is systematically reduced below its SM value, flattening 
the effective scalar potential. The singlet correction 
terms partially compensate this suppression but are 
numerically small due to the constraint $S_{13}^2 < 0.05$.

While our calculation is performed at tree level, 
one-loop corrections to $\lambda_{h_1h_1h_1}$ in the 
NMSSM are known to be of order $\delta\kappa_\lambda 
\sim \pm(3\text{--}8)\%$ and are calculable using 
\texttt{NMSSMCALC}~\cite{Baglio:2014nfa}. Since the 
observed suppression spans a range 
$\Delta\kappa_\lambda \approx 0.10$, which significantly 
exceeds the expected magnitude of loop corrections, 
the qualitative conclusion of universal suppression 
$\kappa_\lambda < 1$ throughout the large-$\lambda$ / 
low-$\tan\beta$ region is expected to be robust. 
We therefore present our results as reliable 
zero-temperature signatures that motivate a forthcoming 
study including full one-loop corrections and an explicit 
finite-temperature analysis via \texttt{CosmoTransitions} 
\cite{Wainwright:2011kj} or \texttt{PhaseTracer} 
\cite{Athron:2020sbe}.

No enhancement of $\kappa_\lambda$ above unity was found for any 
SM-like point, establishing the suppression of the trilinear 
coupling as a \textit{robust prediction} of the scNMSSM in this 
parameter region. The left panel of Figure~\ref{fig:kappa_lambda} 
shows no strong dependence of $\kappa_\lambda$ on $\lambda$, 
consistent with the fact that the suppression is primarily 
determined by the mixing matrix elements $S_{ij}$ rather than by 
$\lambda$ directly. The middle panel shows a mild positive 
correlation between $\kappa_\lambda$ and $\tan\beta$: at fixed 
$\lambda$, increasing $\tan\beta$ reduces the NMSSM contribution 
to $m_{h_1}^2$, requiring a smaller singlet admixture to maintain 
$m_{h_1} \approx 125~\text{GeV}$, which brings $C_V$ and hence 
$\kappa_\lambda$ closer to unity. The right panel confirms that 
$\kappa_\lambda$ increases toward unity as $m_{h_1}$ approaches 
$125.2~\text{GeV}$.

The parameter ranges of the 66 SM-like points are summarized 
in Table~\ref{tab:param_ranges}.

\begin{table}[h]
\caption{Parameter ranges and mean values for the 66 SM-like 
scNMSSM points satisfying $S_{13}^2 < 0.05$ and 
$122~\text{GeV} \leq m_{h_1} \leq 125~\text{GeV}$.}
\label{tab:param_ranges}
\renewcommand{\arraystretch}{1.1}
\centering
\begin{tabular}{|l|l|l|}
\hline
Parameter & Range & Mean \\
\hline
$\lambda$           & $0.520 - 0.641$ & $0.582$ \\
$\kappa$            & $0.341 - 0.498$ & $0.436$ \\
$\tan\beta$         & $2.62  - 4.00$  & $3.57$  \\
$\mu_{\rm eff}$ (GeV) & $105 - 160$   & $131$   \\
$m_{h_1}$ (GeV)    & $122.0 - 125.0$ & $122.9$ \\
$m_{h_2}$ (GeV)    & $131 - 268$     & $197$   \\
$m_{h_3}$ (GeV)    & $383 - 581$     & $483$   \\
$m_{a_1}$ (GeV)    & $8 - 262$       & $119$   \\
$S_{13}^2$          & $0.000 - 0.049$ & $0.018$ \\
$\kappa_\lambda$    & $0.884 - 0.982$ & $0.944$ \\
\hline
\end{tabular}
\end{table}

Table~\ref{tab:benchmark} presents two representative 
benchmark points spanning the full range of $\kappa_\lambda$ 
in our sample, enabling independent verification of our 
results. Point~A, with $\kappa_\lambda = 0.982$, has a 
nearly pure doublet $h_1$ ($S_{13}^2 = 0.003$) and a 
di-Higgs cross section essentially equal to the SM 
prediction. Point~B, with $\kappa_\lambda = 0.884$, 
has the largest singlet admixture in our SM-like sample 
($S_{13}^2 = 0.049$, close to the $S_{13}^2 < 0.05$ 
threshold) and the largest branching ratio 
${\rm BR}(h_3 \to h_1h_1) = 2.57\%$, though the 
resonant cross section $\sigma_{\rm res} = 0.003$ fb 
remains negligible.

\begin{table}[h]
\centering
\caption{Input parameters and key observables for two 
representative benchmark points: Point~A with maximum 
$\kappa_\lambda$ and Point~B with minimum $\kappa_\lambda$ 
among the 66 SM-like scNMSSM points. All masses in GeV, 
cross sections in fb.}
\label{tab:benchmark}
\begin{tabular}{|l|l|l|}
\hline
Quantity & Point~A ($\kappa_\lambda^{\rm max}$) 
         & Point~B ($\kappa_\lambda^{\rm min}$) \\
\hline
\multicolumn{3}{l}{\textit{Input parameters}} \\
\hline
$\lambda$                & 0.6229 & 0.5812 \\
$\kappa$                 & 0.3959 & 0.4029 \\
$\tan\beta$              & 3.632  & 3.923  \\
$\mu_{\rm eff}$ (GeV)    & 144.9  & 145.9  \\
\hline
\multicolumn{3}{l}{\textit{Higgs mass spectrum (GeV)}} \\
\hline
$m_{h_1}$                & 124.3  & 122.3  \\
$m_{h_2}$                & 136.9  & 152.6  \\
$m_{h_3}$                & 523.9  & 580.5  \\
$m_{a_1}$                & 214.9  & 228.6  \\
\hline
\multicolumn{3}{l}{\textit{Mixing matrix elements}} \\
\hline
$S_{11}$                 & 0.2822 & 0.2860 \\
$S_{12}$                 & 0.9578 & 0.9324 \\
$S_{13}$                 & 0.0550 & 0.2208 \\
$S_{13}^2$               & 0.0030 & 0.0488 \\
\hline
\multicolumn{3}{l}{\textit{Trilinear coupling}} \\
\hline
$\kappa_\lambda$         & \textbf{0.982} & \textbf{0.884} \\
\hline
\multicolumn{3}{l}{\textit{Di-Higgs cross sections (fb)}} \\
\hline
$\sigma_{\rm non-res}$   & 33.21  & 32.36  \\
$\sigma_{\rm res}$       & $5\times 10^{-5}$ & $2.6\times 10^{-3}$ \\
${\rm BR}(h_3\to h_1h_1)$ & 0.047\% & 2.571\% \\
\hline
\end{tabular}
\end{table}

\begin{table}[h!]
\centering
\caption{Estimated impact of one-loop corrections on 
$\kappa_\lambda$ for the two benchmark points, based 
on the known NLO correction range 
$\delta\kappa_\lambda \sim \pm(3\text{--}8)\%$ 
from~\cite{Baglio:2014nfa}.}
\label{tab:loop_estimate}
\begin{tabular}{lccc}
\hline\hline
Point & $\kappa_\lambda^{\rm tree}$ 
      & $\kappa_\lambda^{\rm NLO,\,min}$ 
      & $\kappa_\lambda^{\rm NLO,\,max}$ \\
\hline
Point A (max) & 0.982 & 0.904 & 1.061 \\
Point B (min) & 0.884 & 0.813 & 0.955 \\
\hline
\multicolumn{4}{l}{\footnotesize NLO estimates: 
tree-level $\times (1 \pm 0.08)$ for upper/lower bounds.} \\
\hline\hline
\end{tabular}
\end{table}

Table~\ref{tab:loop_estimate} shows that even at the upper end of the estimated NLO correction range, Point~B remains below unity ($\kappa_\lambda^{\rm NLO} \leq 0.955$), while Point~A could in principle cross unity if corrections reach $+8\%$. However, since 
the vast majority of our 66 points have $\kappa_\lambda \leq 0.96$, the qualitative conclusion of universal suppression is expected to be robust against one-loop corrections for most of the parameter space. A definitive assessment requires the full NLO computation via 
\texttt{NMSSMCALC}~\cite{Baglio:2014nfa}, which is deferred to future work.
\subsection{Di-Higgs Production Cross Sections 
at $\sqrt{s} = 13.6~\text{TeV}$}

\begin{figure}[h]
\centering
\includegraphics[width=\textwidth]{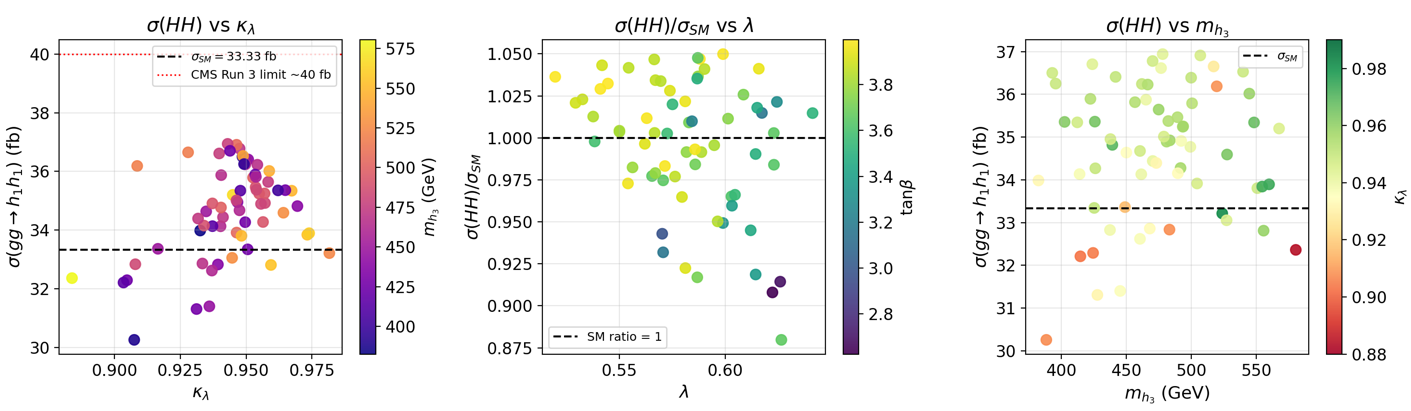}
\caption{Di-Higgs production cross section $\sigma(gg \to h_1h_1)$ at $\sqrt{s} = 13.6$ TeV as a function of (left) $\kappa_\lambda$, (center) $\lambda$, and (right) $m_{h_3}$, for the 66 SM-like points. Color coding indicates $m_{h_3}$ in GeV (left panel), $\tan\beta$ (center panel), and $\kappa_\lambda$ (right panel). The horizontal dashed line shows the SM prediction $\sigma_{\rm SM} = 33.33$ fb. The enhancement of $\sigma(HH)$ above $\sigma_{\rm SM}$ for points with smaller $\kappa_\lambda$ (left panel) reflects the reduction of the destructive interference between the triangle and box diagram amplitudes when $\kappa_\lambda < 1$.}
\label{fig:sigma_HH}
\end{figure}

\begin{figure}[h]
\centering
\includegraphics[width=\textwidth]{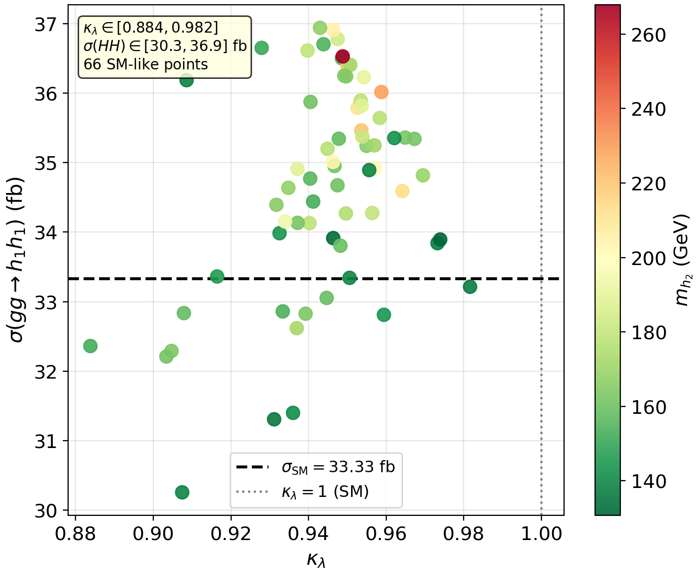}
\caption{Di-Higgs production cross section $\sigma(gg \to h_1h_1)$ at $\sqrt{s} = 13.6$ TeV as a function of $\kappa_\lambda$ for the 66 SM-like scNMSSM points. Color coding indicates $m_{h_2}$ in GeV. The horizontal dashed line shows $\sigma_{\rm SM} = 33.33$ fb and the vertical dotted line marks the SM value $\kappa_\lambda = 1$. The positive correlation between $\sigma(HH)$ and decreasing $\kappa_\lambda$ reflects the reduction of destructive interference between the triangle and box amplitudes when $\kappa_\lambda < 1$.}
\label{fig:kappa_vs_sigma}
\end{figure}

We evaluate both the non-resonant and resonant channels 
contributing to the total di-Higgs production cross section 
$\sigma(gg \to h_1h_1)$ at the LHC. The results are illustrated 
in Figures~\ref{fig:sigma_HH} and~\ref{fig:kappa_vs_sigma}.

\subsubsection*{Non-resonant production}

The non-resonant di-Higgs cross section, dominated by the 
destructive interference between the top-quark box diagram and 
the $h_1$-mediated triangle diagram, is computed using the 
Carvalho et al.\ parametrization~\cite{Carvalho:2015ttv} at 
$\sqrt{s} = 13.6~\text{TeV}$:

\begin{equation}
\frac{\sigma(gg \to h_1h_1)}{\sigma_{\rm SM}} = 
\frac{A_1\kappa_t^4 + A_2\kappa_\lambda^2\kappa_t^2 
    + A_3\kappa_t^4 + A_4\kappa_\lambda\kappa_t^2 
    + A_5\kappa_t^2}
     {A_1 + A_2 + A_3 + A_4 + A_5}
\label{eq:carvalho}
\end{equation}

with coefficients $A_1 = 2.09$, $A_2 = 0.22$, $A_3 = 0.68$, 
$A_4 = -2.12$, $A_5 = 0.96$~\cite{Carvalho:2015ttv}, and the SM 
reference cross section $\sigma_{\rm SM} = 33.33~\text{fb}$ at 
$\sqrt{s} = 13.6~\text{TeV}$ from the LHC Higgs Cross Section 
Working Group~\cite{LHCHiggsCrossSectionWorkingGroup:2016ypw}. The Carvalho et al.\ parametrization was derived assuming the reduced top-quark coupling of $h_1$; $\kappa_t = C_t(h_1) = S_{12}/\sin\beta$ $\kappa_t = 1$. In our sample, $\kappa_t \in [0.97,\,1.01]$, introducing a systematic uncertainty on $\sigma(gg \to h_1h_1)$ of order $\delta\sigma/\sigma \sim 4(\kappa_t - 1) \times (A_1 + A_3 + A_5)/(A_1+A_2+A_3+A_4+A_5) \lesssim 2\%$, 
which is negligible compared to the range of $\sigma/\sigma_{\rm SM} = 0.880$--$1.050$ observed in our scan.

Thus, the cross section is found to lie within the range:

\begin{equation}
\sigma(gg \to h_1h_1) = 30.3 - 36.9~\text{fb}, 
\qquad 
\sigma/\sigma_{\rm SM} = 0.880 - 1.050
\label{eq:sigma_result}
\end{equation}

Of the 66 SM-like points, 35 yield $\sigma > \sigma_{\rm SM}$ 
and 31 yield $\sigma < \sigma_{\rm SM}$. This apparently 
counterintuitive result --- that a suppressed trilinear coupling 
$\kappa_\lambda < 1$ can lead to an enhanced cross section 
$\sigma > \sigma_{\rm SM}$ --- is a direct consequence of the 
destructive interference between the triangle and box diagrams 
in the SM. When $\kappa_\lambda < 1$, the triangle amplitude is 
reduced, which in turn reduces the destructive interference and 
leads to a net enhancement of the total cross section. This effect 
is well-known in the context of anomalous Higgs couplings~\cite{Baglio:2014nfa} 
and is clearly visible in Figure~\ref{fig:kappa_vs_sigma}, where 
$\sigma(HH)$ increases monotonically as $\kappa_\lambda$ decreases 
below unity. This establishes that any detectable deviation in 
di-Higgs production within this model will manifest purely through 
non-resonant kinematic distributions, providing a clean experimental 
signature for future HL-LHC searches.

\subsubsection*{Resonant production via $h_3$}

The resonant contribution from $gg \to h_3 \to h_1h_1$ is 
evaluated as:

\begin{equation}
\sigma_{\rm res}(gg \to h_3 \to h_1h_1) = 
\sigma(gg \to h_3) \times {\rm BR}(h_3 \to h_1h_1)
\label{eq:sigma_res}
\end{equation}

where $\sigma(gg \to h_3) = \sigma_{\rm SM}(m_{h_3}) \times 
C_t(h_3)^2$ and ${\rm BR}(h_3 \to h_1h_1)$ is taken directly 
from the NMSSMTools output.

This channel is found to be entirely negligible for all 66 points, 
with ${\rm BR}(h_3 \to h_1h_1) \leq 2.57\%$ and a maximum 
resonant cross section $\sigma_{\rm res} \leq 0.004~\text{fb}$, 
less than $0.01\%$ of $\sigma_{\rm SM}$. In the scNMSSM under 
the large-$\lambda$ constraint, $h_3$ is predominantly 
singlet-like with $|C_t(h_3)| \leq 0.38$, which severely 
suppresses $\sigma(gg \to h_3)$. Furthermore, the moderate 
branching ratio $h_3 \to h_1h_1$ faces strong competition from 
$h_3 \to t\bar{t}$, $WW$, and $ZZ$ decay channels. This 
establishes that the resonant channel plays no observable role 
in current or near-future collider phenomenology within this 
parameter region.

All 66 predicted cross sections, lying in the range 
$30.3$--$36.9$ fb ($0.88$--$1.05\,\sigma_{\rm SM}$), 
remain well below the current experimental upper limits 
on di-Higgs production. This behaviour is further illustrated in 
Figure~\ref{fig:kappa_vs_sigma}, which shows the monotonic increase of $\sigma(HH)$ as $\kappa_\lambda$ decreases below unity, consistent with the parametrization in Eq.~(\ref{eq:carvalho}): for $\kappa_t \approx 1$, the cross section ratio simplifies to $\sigma/\sigma_{\rm SM} \approx 1 + (A_2\kappa_\lambda^2 + A_4\kappa_\lambda - A_2 - A_4)/(A_1+A_2+A_3+A_4+A_5)$, which is a decreasing function of $\kappa_\lambda$ for $\kappa_\lambda < 1$ given the signs of $A_2$ 
and $A_4$. The recent ATLAS and CMS Run~2 
legacy combination yields an observed (expected) upper 
limit of $2.5\,(1.7)\times\sigma_{\rm SM}$, corresponding 
to approximately 83 (57) fb at 95\% 
CL~\cite{ATLAS-CMS:2025}, while the most recent ATLAS 
Run~3 analysis based on $308~\text{fb}^{-1}$ in the 
$HH \to bb\gamma\gamma$ channel achieves comparable 
sensitivity~\cite{ATLAS:2025}. All predicted cross 
sections in this work therefore remain unconstrained by 
current direct di-Higgs searches. The implications for future $\kappa_\lambda$ measurements are discussed in Section~\ref{sec:Prospects}.

\subsection{Comparison with Previous Work}

Our results can be compared with previous studies of the trilinear 
Higgs self-coupling and di-Higgs production in the NMSSM and 
related models.

In the MSSM, the trilinear self-coupling is suppressed relative 
to the SM value due to radiative corrections from the stop sector, 
but the suppression is typically weak --- $\kappa_\lambda \gtrsim 
0.97$ --- because the MSSM Higgs sector is predominantly 
doublet-like by construction~\cite{Baglio:2014nfa}. Our results in 
the scNMSSM show a comparable but somewhat larger suppression 
($\kappa_\lambda = 0.884$--$0.982$), reflecting the additional 
singlet-doublet mixing present in the NMSSM.

In the general NMSSM with unconstrained parameters, 
Ellwanger~\cite{Ellwanger:2013ova} finds that $\sigma(gg \to hh)$ 
can vary between 0.7 and 2.4 times the SM prediction in broad 
parameter scans, with large enhancements possible when $m_{h_2}$ 
or $m_{h_3}$ lies near the $2m_{h_1}$ threshold. In contrast, 
our scNMSSM results show a much narrower range 
($\sigma/\sigma_{\rm SM} = 0.880$--$1.050$), reflecting the 
additional constraints imposed by the GUT-scale boundary 
conditions and the requirement of SM-like $h_1$ with 
$S_{13}^2 < 0.05$. In particular, the absence of any resonant 
enhancement via $h_3$ in our scan --- despite 
$m_{h_3} \in [383,\,581]~\text{GeV}$ --- is due to the small 
reduced top coupling of $h_3$ ($|C_t(h_3)| \leq 0.38$), which 
strongly suppresses the $gg \to h_3$ production rate.

The most directly comparable study is the recent benchmark 
analysis of Ellwanger et al.~\cite{ellwanger2024benchmark}, which 
provides benchmark lines and planes for Higgs-to-Higgs decays 
in the NMSSM. That study focuses on the general NMSSM without 
GUT-scale boundary conditions and does not systematically 
analyze $\kappa_\lambda$ in the scNMSSM at large-$\lambda$ / 
low-$\tan\beta$. Our work therefore fills a gap in the 
literature by providing the first dedicated analysis of this 
parameter region in the semi-constrained framework, with all 
current LHC Run~3 constraints applied.

Regarding the connection to the EWPT, previous studies of the 
NMSSM phase transition~\cite{Huber:2006wf, Menon:2004wv,Pietroni:1992in} have identified the singlet sector as a key 
driver of a strong first-order EWPT, but have not computed 
$\kappa_\lambda$ for the surviving parameter points or connected 
the phase transition dynamics to the di-Higgs production cross 
section. Our work provides this connection explicitly for the 
scNMSSM at large-$\lambda$ / low-$\tan\beta$, establishing 
$\kappa_\lambda < 1$ as a robust prediction of this parameter 
region.

\section{Implications for the Electroweak Phase Transition}

The cosmological implications of our findings are evaluated within 
the context of the electroweak phase transition (EWPT) and successful 
electroweak baryogenesis (EWBG). A robust EWBG requires a strong 
first-order electroweak phase transition to suppress the sphaleron 
washout mechanism in the broken phase, traditionally parameterized 
by the sphaleron decoupling condition $v_c/T_c \gtrsim 1$, where 
$T_c$ is the critical temperature and $v_c$ is the Higgs vacuum 
expectation value at $T_c$~\cite{Arnold:1987}.

\subsection{Connection Between $\kappa_\lambda$ and the EWPT}

The finite-temperature effective potential governing the EWPT 
can be written schematically as~\cite{Quiros:1999jp}:
\begin{equation}
V_{\rm eff}(\phi, T) = V_0(\phi) + V_{\rm CW}(\phi) + V_T(\phi, T)
\label{eq:Veff}
\end{equation}
where $V_0$ is the tree-level potential, $V_{\rm CW}$ is the 
Coleman-Weinberg one-loop correction, and $V_T$ is the thermal 
contribution. The zero-temperature trilinear Higgs self-coupling 
$\lambda_{h_1h_1h_1}$ is directly related to the third derivative 
of $V_0 + V_{\rm CW}$ evaluated at the electroweak minimum 
$\langle\phi\rangle = v$:
\begin{equation}
\lambda_{h_1h_1h_1} = 
\left.\frac{\partial^3(V_0 + V_{\rm CW})}{\partial\phi^3}
\right|_{\phi=v}
\label{eq:trilinear_potential}
\end{equation}

In the Standard Model, the EWPT is a smooth crossover due to 
the large Higgs boson mass, which fails to satisfy the Sakharov 
conditions for generating the observed baryon asymmetry of the 
Universe~\cite{Kajantie_1996}. A deviation of $\kappa_\lambda$ 
from unity therefore signals a modification of the 
zero-temperature Higgs potential relative to the SM, which 
directly affects the shape of the potential barrier between the 
symmetric and broken phases at finite temperature. This 
connection has been made explicit in several studies of BSM 
scenarios~\cite{Grojean:2005, Espinosa:2007}, which show that 
a strong first-order EWPT generically requires $\kappa_\lambda$ 
to deviate significantly from unity.

Within the scNMSSM under the large-$\lambda$ / low-$\tan\beta$ 
framework investigated in this work, the scalar potential is 
substantially modified. The universal suppression $\kappa_\lambda \in 
[0.884,\,0.982]$ found in our scan is qualitatively 
consistent with the conditions favouring a modified 
Higgs potential that could support a strengthened 
first-order transition, though a quantitative 
determination of the phase transition strength requires 
an explicit computation of $V_{\rm eff}(\phi, T)$ 
and the ratio $\xi_c = v_c/T_c$, which is deferred 
to future work. In the SM, the trilinear coupling fixes the curvature of the potential near $v$; its suppression below the SM value reflects a flattening of 
the potential in this region, which is a necessary (though not 
sufficient) condition for the existence of a barrier separating 
the two phases at finite temperature~\cite{Grojean:2005, 
Espinosa:2007}.

\subsection{The scNMSSM and the First-Order EWPT}

In the NMSSM, the electroweak phase transition can be strongly 
first order in specific regions of parameter space where the 
singlet sector contributes to the potential 
barrier~\cite{Huber:2006wf, Menon:2004wv}. The relevant 
contributions arise from two sources.

\textbf{Tree-level barrier.} At large $\lambda$, the interaction 
between the electroweak doublets and the singlet field $S$ is 
enhanced. As the singlet acquires its vacuum expectation value 
$v_s = \mu_{\rm eff}/\lambda$, the singlet-doublet interaction 
$\lambda S H_u H_d$ in the superpotential generates a tree-level 
cubic term in the effective potential:
\begin{equation}
\Delta V_{\rm tree} \supset 
-\frac{\lambda\,\mu_{\rm eff}}{\sqrt{2}}\,\phi^3\sin 2\beta 
+ \ldots
\label{eq:tree_barrier}
\end{equation}
This cubic term creates a barrier between the symmetric and 
broken phases even at zero temperature, which is strengthened 
at finite temperature and can drive a first-order transition. 
The strength of this barrier is controlled by $\lambda\mu_{\rm eff}$, 
which in our scan lies in the range 
$\lambda\mu_{\rm eff} \in [55,\,102]~\text{GeV}$.

The suppression of $\kappa_\lambda$ observed in our collider 
analysis is deeply interconnected with this tree-level barrier 
mechanism. The doublet-singlet mixing that reduces $C_V$ below 
unity --- and hence suppresses $\lambda_{h_1h_1h_1}$ relative 
to its SM value --- is the same mixing that generates the cubic 
term in Eq.~(\ref{eq:tree_barrier}). A larger $|S_{13}|$ 
generally implies both a stronger modification of 
$\lambda_{h_1h_1h_1}$ and a stronger potential barrier, though 
the relationship is non-trivial due to the interplay with 
$\kappa$ and $\mu_{\rm eff}$.

\textbf{Loop-induced barrier.} The soft SUSY-breaking trilinear 
terms $A_\lambda$ and $A_\kappa$, together with the $\kappa S^3/3$ 
self-coupling, contribute additional loop-induced terms to the 
thermal potential that can enhance the strength of the 
transition~\cite{Pietroni:1992in, Carena:1996wj}. In the scNMSSM 
with GUT-scale boundary conditions, these contributions are 
constrained by the requirement of correct electroweak symmetry 
breaking, but remain non-negligible.

This tree-level driven mechanism contrasts sharply with the 
radiative contributions typical of the MSSM or the SM with low 
cutoffs, where heavy stop loops or exotic particles are required 
to induce a first-order transition. By relying on the tree-level 
coupling $\lambda$, the scNMSSM naturally circumvents severe 
fine-tuning constraints while satisfying the necessary geometric 
conditions for a strong phase transition.

Our finding that $\kappa_\lambda \in [0.884,\,0.982]$ 
with mean $\langle\kappa_\lambda\rangle = 0.944 \pm 0.018$ 
indicates that the scNMSSM at large-$\lambda$ / 
low-$\tan\beta$ systematically modifies the Higgs 
potential in a direction qualitatively consistent with 
the requirements of a strengthened first-order EWPT, 
though the magnitude of the modification is modest 
and a definitive conclusion requires a dedicated 
finite-temperature analysis. While a full 
quantitative determination requires computing 
$V_{\rm eff}(\phi, T)$ explicitly and evaluating $v_c/T_c$ --- 
using dedicated tools such as \texttt{CosmoTransitions}~\cite{
Wainwright:2011kj} or \texttt{PhaseTracer}~\cite{Athron:2020sbe} --- 
this is deferred to future work. Nevertheless, the persistent 
suppression of $\kappa_\lambda$ across all 66 
phenomenologically viable points provides a compelling, 
model-independent collider footprint of a framework capable 
of accommodating successful electroweak baryogenesis.

\subsection{Observational Prospects} \label{sec:Prospects}


The connection between $\kappa_\lambda$ and the EWPT has 
important implications for the experimental program at current 
and future colliders.

\textbf{LHC and HL-LHC.} The level of suppression found in our scan, 
$\Delta\kappa_\lambda = \kappa_\lambda - 1 \in 
[-0.116,\,-0.018]$, is beyond the reach of current 
di-Higgs searches. The most stringent experimental 
constraint to date comes from the ATLAS+CMS Run~2 
legacy combination~\cite{ATLAS-CMS:2025}, which 
constrains $\kappa_\lambda \in [-0.4,\,6.3]$ at 95\% 
CL --- far too broad to resolve the $\sim$2--12\% 
suppression predicted in this work. The HL-LHC with 
$3000~\text{fb}^{-1}$ is projected to constrain $\kappa_\lambda$ 
to approximately $\kappa_\lambda \in [0.5,\,1.5]$ at 68\% CL 
from di-Higgs production alone~\cite{Cepeda:2019klc}, which would 
still not be sufficient to individually resolve the $\sim$2--12\% 
suppression predicted by our scan. However, a global combination 
of single-Higgs and di-Higgs measurements at the HL-LHC could 
improve the sensitivity to $\Delta\kappa_\lambda \sim \pm 0.2$ 
\cite{deBlas:2019rxi}, which would begin to probe the upper end of 
our predicted range.

\textbf{Future lepton colliders.} The most promising prospects 
come from future lepton colliders. The International Linear 
Collider (ILC) at $\sqrt{s} = 500~\text{GeV}$ with 
$4000~\text{fb}^{-1}$ is projected to measure $\kappa_\lambda$ 
with a precision of $\Delta\kappa_\lambda \sim \pm 0.1$ 
\cite{Bambade:2019fyw}, sufficient to probe $\kappa_\lambda$ values 
as small as 0.9. The Compact Linear Collider (CLIC) at 
$\sqrt{s} = 3~\text{TeV}$ could achieve 
$\Delta\kappa_\lambda \sim \pm 0.05$~\cite{Roloff_2020}, 
sensitive to the full range predicted by our scan. At a muon 
collider operating at $\sqrt{s} = 10~\text{TeV}$, the 
sensitivity could reach $\Delta\kappa_\lambda \sim \pm 0.02$ 
\cite{Han:2020pif}, providing a definitive test of the scNMSSM 
predictions.

The complementarity between the di-Higgs measurement of 
$\kappa_\lambda$ and direct searches for the heavy scalars 
$h_2$ and $h_3$ should also be noted. In our scan, 
$m_{h_2} \in [131,\,268]~\text{GeV}$ and 
$m_{h_3} \in [383,\,581]~\text{GeV}$. The $h_2$ state is 
potentially accessible at the LHC through 
$gg \to h_2 \to WW,\,ZZ,\,\tau\tau$ channels, while $h_3$ 
could be searched for in $gg \to h_3 \to h_1h_1,\,h_1h_2,\,
t\bar{t}$ final states. A simultaneous observation of a 
suppressed $\kappa_\lambda$, a light $h_2$, and a heavy $h_3$ 
would provide strong evidence for the scNMSSM at large-$\lambda$ 
/ low-$\tan\beta$ and its associated modification of the Higgs potential relevant for electroweak baryogenesis. 

\section{Conclusions and Future Outlook}

We have presented a systematic analysis of the trilinear Higgs self-coupling $\kappa_\lambda$ and the di-Higgs production cross section $\sigma(gg \to h_1h_1)$ in 
the semi-constrained NMSSM (scNMSSM) at large-$\lambda$ / 
low-$\tan\beta$, incorporating GUT-scale universal boundary 
conditions, full two-loop Higgs mass precision, and current 
LHC Run 3 constraints. Our scan was strategically targeted at 
this well-motivated regime, where the NMSSM tree-level 
contribution $\Delta m_h^2 \propto \lambda^2 v^2 \sin^2 2\beta$ 
is maximized. Out of one million initial parameter points, a 
robust subset of 66 SM-like points successfully survived all 
combined filters, defined by $122~\text{GeV} \leq m_{h_1} \leq 
128~\text{GeV}$ and $S_{13}^2 < 0.05$.

Our analysis demonstrates a \textbf{universal suppression} of 
the trilinear Higgs self-coupling ratio across all viable points. 
The key quantitative predictions are:

\begin{equation*}
\kappa_\lambda = 0.884\text{--}0.982, \quad 
\langle\kappa_\lambda\rangle = 0.944 \pm 0.018
\end{equation*}

\begin{equation*}
\sigma(gg \to h_1h_1) = 30.3\text{--}36.9~\text{fb}, \quad
\sigma/\sigma_{\rm SM} = 0.880\text{--}1.050
\end{equation*}

No enhancement of $\kappa_\lambda$ above unity was found for any 
SM-like point, establishing this suppression as a robust 
prediction of the scNMSSM in this parameter region. The 
persistent modification is inherently driven by doublet-singlet 
scalar mixing at tree level: the reduced doublet fraction 
$C_V < 1$ systematically suppresses the dominant $C_V^3$ term 
in $\lambda_{h_1h_1h_1}$. The non-resonant cross section 
exceeds $\sigma_{\rm SM}$ for 35 of the 66 points and falls 
below it for the remaining 31, reflecting the reduced destructive 
interference between the triangle and box amplitudes when 
$\kappa_\lambda < 1$. The resonant contribution via 
$gg \to h_3 \to h_1h_1$, with $m_{h_3} \in [383,\,581]~\text{GeV}$, 
is entirely suppressed ($\leq 0.004~\text{fb}$) due to the 
predominantly singlet-like nature of $h_3$ and its small reduced 
top coupling $|C_t(h_3)| \leq 0.38$. This establishes that any 
detectable deviation in di-Higgs production within this model 
will manifest purely through non-resonant kinematic distributions.

We discussed these collider phenomenology findings in the 
context of the cosmological electroweak phase transition. The 
universal suppression of $\kappa_\lambda$ reflects a distinctive 
zero-temperature flattening of the scalar potential, arising from 
the same singlet-doublet mixing that generates a tree-level cubic 
barrier in the effective potential with strength 
$\lambda\mu_{\rm eff} \in [55,\,102]~\text{GeV}$. This mechanism 
naturally circumvents the severe fine-tuning typical of 
radiative loop-driven phase transitions in the MSSM, providing 
qualitative evidence that the large-$\lambda$ / low-$\tan\beta$ 
region of the scNMSSM can accommodate the conditions required 
for successful electroweak baryogenesis.

Regarding experimental prospects, the predicted 
$\Delta\kappa_\lambda \in [-0.116,\,-0.018]$ is beyond the 
reach of current LHC Run~3 di-Higgs searches but lies within 
the projected reach of the HL-LHC global combination 
($\Delta\kappa_\lambda \sim \pm 0.2$) and is fully accessible 
at future lepton colliders: ILC ($\pm 0.1$), CLIC ($\pm 0.05$), 
and a 10~TeV muon collider ($\pm 0.02$). This distinct 
non-resonant collider footprint provides an exciting benchmark 
for future high-energy collider programs. Several natural 
extensions of this work are identified:

\begin{enumerate}
    \item The inclusion of one-loop corrections to 
    $\lambda_{h_1h_1h_1}$ via NMSSMCALC~\cite{Baglio:2014nfa, Ender_2012}, 
    expected to shift $\kappa_\lambda$ by a few percent without 
    altering the qualitative conclusion of universal suppression.
    
    \item An explicit computation of the finite-temperature 
    effective potential $V_{\rm eff}(\phi, T)$ and the 
    sphaleron decoupling ratio $\xi_c = v_c/T_c$ for the 
    surviving points, using \texttt{CosmoTransitions} 
    \cite{Wainwright:2011kj} or \texttt{PhaseTracer} 
    \cite{Athron:2020sbe}, to quantitatively establish which 
    points satisfy the baryogenesis condition 
    $\xi_c \gtrsim 1$.
    
    \item An extension to the CP-violating NMSSM, where 
    additional sources of CP violation beyond the CKM matrix 
    could contribute to the observed baryon asymmetry.
    
    \item The computation of di-Higgs signal rates in specific 
    final states --- $b\bar{b}\gamma\gamma$, 
    $b\bar{b}\tau^+\tau^-$, and $b\bar{b}b\bar{b}$ --- 
    including detector simulation, to provide concrete 
    predictions for HL-LHC searches.
\end{enumerate}

The zero-temperature signatures established in this study 
offer a compelling and model-independent experimental handle 
to probe the cosmological viability of the scNMSSM. A 
simultaneous observation of suppressed $\kappa_\lambda$, a 
moderate $h_2$ in the range $[131,\,268]~\text{GeV}$, and 
a heavy singlet-like $h_3$ in the range 
$[383,\,581]~\text{GeV}$ would constitute a distinctive 
multi-messenger signature of this model, accessible to 
the comprehensive experimental program at the LHC and 
future colliders.

\end{document}